# Universal Mechanism for Quantitative Understanding of Global Ozone Depletion

Qing-Bin Lu*

Department of Physics and Astronomy and Departments of Biology and Chemistry, University of Waterloo, 200 University Avenue West, Waterloo, Ontario, Canada

*Corresponding author (Email: qblu@uwaterloo.ca)

**Abstract**: This paper formulates the cosmic-ray(CR)-driven electron-induced reaction (CRE) mechanism to provide a quantitative understanding of global ozone depletion. Based on a proposed electrostatic bonding mechanism for charged-induced adsorption of molecules on surfaces and on the measured dissociative electron transfer (DET) cross sections of ozone depletion substances (ODSs) adsorbed on ice, an analytical equation is derived to give atmospheric chlorine atom concentration:

$$[Cl] = \sum_i k^i \theta_{ODS}^i \Phi_e^2,$$

where $\Phi_e$ is the CR-produced prehydrated electron ($e_{pre}^-$) flux on atmospheric particle surfaces, $\theta_{ODS}^i$ is the surface coverage of an ODS, and $k^i$ is the ODS's effective DET coefficient comprising the DET cross section, lifetimes of surface-trapped $e_{pre}^-$ and $Cl^-$, and particle surface area density. With concentrations of ODSs as the sole variable, our calculated results of time-series ozone depletion rates in global regions in the 1960s, 1980s and 2000s show generally good agreement with observations, particularly with ground-based ozonesonde data and satellite-measured data over Antarctica and with satellite data in the tropics in a narrow altitude band at 13-20 km. Good agreements with satellite data in the Arctic and midlatitudes are also found. A new insight into the denitrification effect on ozone loss is given quantitatively. But this equation overestimates tropospheric ozone loss at northern midlatitudes and the Arctic, likely due to increased ozone production by the halogen chemistry in polluted regions. Finally, ozone maps from ozonesonde data clearly reveal the scope of the tropical ozone hole. The results render confidence in applying the CRE equation to achieve a quantitative understanding of global ozone depletion.

**Introduction**

In a recent paper (*1*), the author used mainly the ground-based Trajectory-mapped Ozonesonde dataset for the Stratosphere and Troposphere (TOST), which is a global 3D (i.e. latitude, longitude, altitude) climatology of tropospheric and stratospheric ozone derived from the World Ozone and Ultraviolet Radiation Data Centre (WOUDC) global ozone sounding record by trajectory mapping (*2, 3*), together with satellite-measured stratospheric temperature climatology as well as other ozone and ozone-depletion substance (ODS) datasets, to reveal the largest ozone hole over the tropics (30°S-30°N). The latter is defined as an area with $O_3$ loss ≥ 25% relative to the undisturbed $O_3$ layer when there were no significant stratospheric chlorofluorocarbons (CFCs, major ODSs), approximately in the 1960s. It worth noting that it is the observed data alone that revealed the largest tropical $O_3$ hole, *independent of any physical model, either photochemical models or the cosmic-ray driven electron-induced reaction (CRE) model* (*1*).

In the Response to a Comment by Chipperfield et al. on the finding of the tropical ozone hole, the author (*4*) further presented the WOUDC's TOST, NASA's GOZCARDS satellite and NOAA's BDBP data sets in both their raw/source and processed forms. All the data from the three independent data sets in both raw and processed forms have clearly confirmed large ozone depletion up to 80% in the tropics across the seasons since the 1980s, consistent with the original finding. The findings reported in the original paper (*1*) and the Response (*4*) call for more careful studies of tropical ozone depletion.

Interestingly, Randel et al. (*5*) in 1999 gave an excellent review on estimates of trends in the vertical profile of ozone from analyses of satellite, ground-based, and balloon measurements for the period 1979–1998. For northern midlatitudes, combined trend estimates for the period 1979–96 showed statistically significant negative trends at all altitudes between 10 and 45 km, with two local maxima: 7.4±2.0% per decade at 40 km and 7.3±4.6% per decade at 15 km altitude in the 1980s and 1990s. In the upper stratosphere, the observed ozone trends have good quantitative agreement with model calculations of ozone loss due to anthropogenic chlorine emissions through the photochemical mechanism, indicating good understanding of ozone depletion in this altitude range. In the lower stratosphere, long-term ozone negative trends show reasonable agreement between SAGE satellite data and ozonesonde data over 15–27 km; the major contribution to column ozone loss occurred over altitudes 10–25 km. Of particular interest are their analyses of SAGE II satellite data for the period 1984–98, which exhibited negative trends over much of the globe in the lower stratosphere, *with largest percentage changes (on the order of 10% per decade) in the tropics below 20 km*. Randel et al. (*5*) noted that such tropical $O_3$ depletion was *not* predicted by photochemical model calculations. They also commented that satellite measurements in the tropics were sensitive to details of the retrieval algorithm and independent estimates of tropical $O_3$ trends from ozonesonde measurements lacked. Therefore, they concluded that the possibility of large $O_3$ loss in the tropical lower stratosphere remained a topic of continued research (*5*).

It is worthwhile to note that the satellite data set of SAGE I/II, which continued the SAGE measurements of stratospheric ozone from 1979-2005, has the longest period of satellite measurements during the drastic ozone depletion in the 1980s and 1990s and has been extensively validated, proving very valuable in determining ozone trends (*6, 7*). All long-term merged satellite data sets rely mostly on SAGE I/II data for pre-1997 periods, as reviewed in the recent SPARC's LOTUS Report (*7*). It is the long-term, stable SAGE II data set that led to the first observation of the largest ozone percentage decreases in the tropical lower stratosphere by Randel et al. (*5*). The observed results in the recent paper (*1*) and the Response (*4*) seem to indicate that the TOST ozone sonde dataset (*3*) may have provided the very exact needed dataset for independent estimates. As noted by Liu et al. (*3*), the TOST ozone climatology was built with ozone data obtained from an independent source and covered a record since 1965 that is much longer than the satellite record and most crucially, the TOST dataset depends on neither a priori data set nor photochemical modeling and thus provides independent information and insights that can supplement satellite data and model simulations.

It has been over two decades since the publication of the review by Randel et al (*5*). However, the status of ozone research remains essentially unchanged: Chemistry-climate models (CCMs), considering the chemical effects of ODSs on $O_3$, were able to reproduce the observed $O_3$ loss of northern mid-latitude $O_3$ in the middle and upper stratosphere, but large discrepancies between CCMs and observations in the lower stratosphere still exist (*7, 8*). Moreover, the uncertainties in $O_3$ profile for both models and ground- and satellite-based observations of pre-1997 $O_3$ trends are very large (up to 20% $O_3$ loss per decade) in the lowermost stratosphere (see, e.g., Figs. 5.9 and 5.10 in the recent LOTUS Report (*7*)); researchers' confidence in trend results is reduced in the lower stratosphere due to large natural variability, low $O_3$ values, and decreased sensitivity of satellite observations. Additional and independent research is called to put the trend results in the lower stratosphere on a more solid ground (*7*).

As noted by Randel et al. (*5*), the vertical profiles of $O_3$ trends can provide a fingerprint for the mechanisms of $O_3$ depletion. Therefore, it is crucial to obtain quantitative estimates of vertical profiles of $O_3$ loss and to compare them with observations. This article aims to derive an analytical quantification equation with no fitting parameters of global $O_3$ depletion based on the CRE mechanism, which comprises two major cosmic ray (CR)-driven processes, the charge-induced adsorption of ODSs on surfaces of atmospheric particles and the dissociative electron transfer (DET) reaction of the thus adsorbed ODSs on the surfaces under CR radiation. For the former process, we will apply and modify a universal electrostatic bonding mechanism proposed by van Driel and co-workers (*9, 10*) to find the coverages of ODSs adsorbed on the CR-caused charged surfaces of atmospheric particles. For the latter process, the DET reactions of ODSs adsorbed on ice surfaces or water clusters have been well studied both experimentally by Lu, Madey and Sanche (*11-17*) and Wolf and co-workers (*18, 19*) and theoretically by Fabrikant (*20, 21*). Our theoretical results will then be compared with the observed vertical profiles of $O_3$ loss from both ozonesonde and satellite measurements. Finally, global maps of $O_3$ depletion, obtained from the TOST dataset, will be presented to confirm the scope of the tropical $O_3$ hole. This study mainly focuses on global $O_3$ depletion in the stratosphere, while it may also have applicability to understanding $O_3$ loss in the troposphere, especially in less polluted regions.

**Ozone Depletion Theories**

# I. Chlorofluorocarbons (CFCs) and heterogeneous chemical reactions

In 1974, Molina and Rowland (*22*) put forward a hypothesis that Cl atomic radicals released from the photolysis of chlorofluorocarbons (CFCs) in the upper stratosphere at altitudes of 30-50 km can participate in the Cl catalytic chain reaction proposed by Stolarski and Cicerone (*23*) for stratospheric ozone depletion

$$Cl + O_3 \rightarrow ClO + O_2 \quad (1a)$$
$$ClO + O \rightarrow Cl + O_2 \quad (1b)$$
$$\text{Net: } O_3 + O \rightarrow O_2 + O_2$$

This reaction chain has a close analogy with the NO free radical catalytic chain reaction proposed by Crutzen (*24*). It is now known that this catalytic cycle can lead to removal of about $10^5$ ozone molecules per chlorine atom (*25*).

The unexpected discovery by Farman et al. (*26*) in 1985 of the Antarctic ozone hole in the lower stratosphere below 20 km was neither predicted nor explainable by the originally proposed photolysis of CFCs. To find a chemical mechanism for the Antarctic $O_3$ hole, a mixed photochemical mechanism consisting of four major processes was proposed. (1) The solar UV photolysis of CFCs produces Cl and ClO radicals that then react with other atmospheric molecules ($CH_4$ and $NO_2$) to generate inorganic chlorine species (HCl and $ClONO_2$) in the tropical upper stratosphere at the altitudes of 30-40 km. (2) The stable reservoirs HCl and $ClONO_2$ are then transported to the polar lower stratosphere via the Brewer-Dobson circulation (BDC). (3) In the dark winter with the absence of sunlight, polar air cools and descends; the cold polar vortex isolates a continent-size body of air, in which polar stratospheric clouds (PSCs) consisting of water ice or nitric acid/ice particles are formed. HCl is adsorbed onto the surfaces of PSC particles and subsequently collisions with gaseous $ClONO_2$ and $N_2O_5$ release photoactive species $Cl_2$, HOCl and $ClNO_2$ into the gas phase and build up $HNO_3$ in PSC particles (*27-29*):

$$ClONO_2(g) + HCl(s) \rightarrow Cl_2(g) + HNO_3(s) \quad (2a)$$
$$N_2O_5(g) + HCl(s) \rightarrow ClNO_2(g) + HNO_3(s) \quad (2b)$$
$$N_2O_5(g) + H_2O(s) \rightarrow 2HNO_3(s) \quad (2c)$$
$$ClONO_2(g) + H_2O(s) \rightarrow HOCl(g) + HNO_3(s) \quad (2d)$$
$$HOCl(g) + HCl(s) \rightarrow Cl_2(g) + H_2O(s) \quad (2e)$$

(s, solid; g, gas). Gas-phase $Cl_2$, HOCl and $ClNO_2$ resulting from the above heterogeneous chemical reactions are accumulated in the polar vortex in the winter. Solomon et al. (*27*) first suggested that the reaction in 2a is the most important one for halogen activation, while Crutzen and Arnold (*30*) and McElroy et al. (*31*) suggested that the 'denoxification' converting NO and NO$_2$ into less reactive HNO$_3$ helps to maintain high levels of active chlorine in the polar stratosphere. (4) With the return of sunlight in the spring, Cl$_2$, ClNO$_2$ and HOCl are rapidly photolyzed to yield Cl atoms that destroy O$_3$ via the Cl catalytic cycles

$$Cl_2(g) + h\nu \rightarrow 2Cl \qquad (3a)$$
$$ClNO_2(g) + h\nu \rightarrow Cl + NO_2 \qquad (3b)$$
$$HOCl(g) + h\nu \rightarrow Cl + OH \qquad (3c)$$

To explain the formation of the Antarctic ozone hole, Molina and Molina (*32*) suggested that a ClO dimer (Cl$_2$O$_2$) is involved and its subsequent photolysis regenerates Cl atoms, which then destroy O$_3$ via the Cl-catalytic cycle

$$2(Cl + O_3) \rightarrow 2(ClO + O_2) \qquad (4a)$$
$$ClO + ClO + M \rightarrow Cl_2O_2 + M \qquad (4b)$$
$$Cl_2O_2 + h\nu \rightarrow Cl + ClOO \qquad (4c)$$
$$ClOO + M \rightarrow Cl + O_2 + M \qquad (4d)$$

Net: $2\,O_3 \rightarrow 3\,O_2$

In addition, McElroy et al. (*31*) also suggested the coupling between ClO and BrO in O$_3$ destruction through the reaction cycle:

$$Cl + O_3 \rightarrow ClO + O_2 \qquad (5a)$$
$$Br + O_3 \rightarrow BrO + O_2 \qquad (5b)$$
$$ClO + BrO \rightarrow Cl + Br + O_2 \qquad (5c)$$

Net: $2\,O_3 \rightarrow 3\,O_2$

In the above catalytic cycle proposed by Molina and Molina, the photolysis of Cl$_2$O$_2$ is rapid to yield two Cl atoms and is not the rate-limiting step. Thus, the observed O$_3$ depletion rate is often related to the measured ClO and BrO concentrations by the following expression (*33-35*):

$$-(d[O_3]/dt) (\equiv d[Cl]_{4a}/dt + d[Cl]_{5a}/dt + d[Br]_{5b}/dt) = 2(k_1[ClO]^2 + k_2[ClO][BrO]) \qquad (6)$$

where $k_1$ and $k_2$ are the rate constants for the ClO dimer formation (reaction 4b) and the coupling between ClO and BrO (reaction 5c) respectively. The first term is dominant as the [ClO] concentration is about 2 orders of magnitude higher than the [BrO] concentration.

In essence, O$_3$ is directly depleted by the reactive halogen atom via reaction (1a/4a) or reaction (5a/5b) in any of the above catalytic cycles. Thus, the O$_3$ loss rate is more directly expressed in the dependence on Cl and Br atomic concentrations [Cl] and [Br] (*36*)

$$-(d[O_3]/dt) \equiv (k_{Cl}[\text{Cl}] + k_{Br}[\text{Br}])[O_3] \qquad (7)$$

where $k_{Cl}$ ($k_{Br}$) is the rate constant for the reaction of Cl (Br) atom with ozone, with $k_{Cl}$ =$2.9\times10^{-11}\exp(-260/T)$ cm$^3$s$^{-1}$ (35). Eq. 7 is always equivalent to Eq. 6 in reaction kinetics of the catalytic cycles, since we have relationships $2k_1[\text{ClO}]^2 = k_{Cl}[\text{Cl}]_{4a}[O_3]$ and $2k_2[\text{ClO}][\text{BrO}] = (k_{Cl}[\text{Cl}]_{5a} + k_{Br}[\text{Br}]_{5b})[O_3]$ with $[\text{Cl}] = [\text{Cl}]_{4a} + [\text{Cl}]_{5a}$ and $[\text{Br}]= [\text{Br}]_{5b}$ (see an analogous equation $k_{1b}[\text{O}][\text{ClO}] = k_{Cl}[\text{Cl}][O_3]$, where $k_{1b}$ is the rate constant for reaction (1b), in the catalytic cycle proposed by Stolarski and Cicerone (23)).

For quantitative estimates of ozone depletion based on the field measurements of [ClO] concentrations, the expression in Eq. 6 is preferred to Eq. 7 as the atomic concentration [Cl] is orders of magnitude smaller than [ClO] and hence [Cl] is more difficult to reliably measure. Moreover, the halogen chemistry in the troposphere may be far more complex in that halogen atomic radicals may involve in the processes leading to not only ozone loss but also increased ozone production, especially in polluted regions (37-39). Nevertheless, Eq. 7 has been used to study ozone depletion during the 'polar sunrise', which is referred to a dramatic loss of surface level ozone measured in the Arctic during springtime, when the atmosphere is first exposed to sunlight after several month darkness. This destruction of tropospheric ozone is correlated with high average diurnal Cl and Br atom concentrations in polar regions (36) and mid-latitudes (40) derived from measurement of the relative loss rates of a group of selected alkanes. Integration of Eq. 7 over a 24-hour period yields the amount of O$_3$ removed in each diurnal cycle (36). For stratospheric halogen chemistry, Eq. 7 is generally valid to represent the change rate of ozone and is convenient to obtain the rate of ozone loss in percentage without the necessity of being given the absolute O$_3$ concentration [O$_3$], provided that the processes leading to the formation of Cl (Br) atomic radicals are well understood. Another advantage of Eq. 7 is its absence of the coupling term between ClO and BrO radicals that exists in Eq. 6. Moreover, the atmospheric abundances of human-made CFCs and HCFCs have been dominant in all ODSs, while the concentration of the Br-containing ODS (mainly CH$_3$Br), which has a main natural origin, has had no major changes since the 1960s. Also, the rate constant $k_{Cl}$ is 10 times larger than $k_{Br}$ at 298 K (35, 36). All these factors lead to a good approximation for quantitative estimates of *time-series changes in anthropogenic ozone loss rate*

$$-(d[O_3]/dt) \approx k_{Cl}[\text{Cl}][O_3] \qquad (8)$$

This equation will be used for this current study, in which [Cl] will be theoretically derived.

II. **The CRE mechanism**

The dissociative electron attachment (DEA) in the gas phase and the dissociative electron transfer (DET) reaction for adsorption on ice surfaces or water clusters or in liquid water of halogenated molecules have been well studied (20, 21, 41-45)

$$\text{AB} + e^- (\sim 0 \text{ eV}) \rightarrow \text{AB}^{*-} \rightarrow \text{A}^- + \text{B} \qquad \text{(DEA)} \qquad (9)$$

$$AB + e_{pre}^{-}(H_2O)_n \rightarrow AB^{*-} \rightarrow A^{-} + B \quad \text{(DET)} \quad (10)$$

where the transient species $AB^{*-}$ is a vibrationally-excited state of the molecule and $e^{-}/e_{pre}^{-}$ is a free electron or a prehydrated electron trapped in ice or water. The DEAs of most halogen(Cl, Br and I)-containing molecules (such as CFCs) are exothermic, and therefore DEAs of these molecules can effectively occur at nearly zero eV electrons in the gas phase. Remarkably, the giant enhancements in DEA of CFCs adsorbed on ice surfaces was observed by Lu, Madey and Sanche (*11-15*). This reaction is called DET as it involves the trapping and transfer of an $e_{pre}^{-}$ on ice surface. The exothermic energies of the DEA reactions on $H_2O$ ice or in liquid water are enhanced by 1–2 eV due to the effect of the polarization potential (*46*). This leads to strong resonances of anion states of Cl-, Br- and I-containing molecules with $e_{pre}^{-}$ that is weakly-bound at −1.5 to −1.0 eV (*16, 47*). Thus, highly effective resonant DET can occur for organic and inorganic Cl-, Br- and I-containing molecules on $H_2O$ ice or in liquid water (*16, 17*). At the temperature of about 25 K, the DET cross sections were measured to be $(1.0$-$1.3) \times 10^{-14}$ and $\sim 8.9 \times 10^{-14}$ $cm^2$ for $CF_2Cl_2$ and $CFCl_3$ adsorbed on $H_2O$ ice respectively, $\sim 6.0 \times 10^{-12}$ $cm^2$ for $CF_2Cl_2$ on $NH_3$ ice, and $\sim 4.0 \times 10^{-15}$ $cm^2$ for HCl on $H_2O$ ice, which are $10^6$–$10^8$ times their photodissociation cross sections (*11-15*). The highly effectively DET reaction of CFCs was later confirmed by Kim and co-workers (*48*) and Wolf and co-workers (*18, 19*), using real-time femtosecond laser spectroscopic methods.

The temperature dependences of DEA and DET reaction rates were studied previously by Fabrikant and co-workers (*49*) and us (*50*). Rate coefficients $k(T)$ for DEA and DET reactions of many molecules generally exhibit a rise with increasing temperature $T$, which is often represented by an Arrhenius or Arrhenius-like equation $k(T) \propto \exp[-E_a/(k_BT)]$ with the activation energy $E_a$ deduced from fits to the experimental data $k(T)$. In a direct DEA process of halogenated ODSs including $CCl_4$, CFCs and HCFCs, this behavior indicates an energy barrier of 0-0.6 eV for the anion on its path to the dissociated products, which is associated with Franck–Condon factors for getting from the initial neutral vibrational levels to the dissociating final anion state (*49*). In a DET reaction, this energy barrier is much smaller and associated with the structural reorganization energy (0-0.1 eV) of the medium (solid ice or liquid water) (*50*). In the latter case, the electron transfer occurs by a harpooning reaction on the surface. In this study, we will simply use $E_a$=0.05 eV for all ODSs and the DET cross sections measured at 25 K to extend the DET cross sections to various temperatures in the stratosphere and troposphere.

In the stratosphere and troposphere, the major source of electrons is simply the atmospheric ionization caused by cosmic rays (CRs) originating from deep space. Although the measured DEA cross sections of gas-phase CFCs to low-energy free electrons near zero eV are very large, about $10^4$ times the photodissociation cross sections of CFCs (*42, 43*), and it was suggested by Peyerimhoff and co-workers (*44, 45*) that the DEA of CFCs must be seen in competition to the photodissociation process and must be considered as a factor in evaluating stratospheric $O_3$ depletion, the gaseous DEA process was thought not to be significant for stratospheric ozone depletion due to the very low free electron density (*51*). However, it was also noted that the understanding of stratospheric negative-ion chemistry was rather speculative (*52, 53*). Then, the surprizing observation of giant enhancements in DET of CFCs adsorbed on ice surfaces has stimulated continued studies. The DEA/DET reactions of halogenated molecules can generate reactive neutral radicals and halogen anions (*11, 17*), and most of the halogen anions are trapped at ice surfaces due to the image potential (*11-17, 19, 48*). It has been known for some time that ionic (heterogeneous) reactions on surfaces can effectively convert $Cl^-/Br^-$ into photoactive $XNO_2$, $X_2$, and HOX (X=Cl, Br), as demonstrated on solid or aqueous NaCl or sea-salt aerosols by Finlayson Pitts and co-workers (*54-56*). A known effective conversion of a $Cl^-$ ion into an active chlorine species is the reaction with the nocturnal $NO_X$ species, $N_2O_5$

$$N_2O_5(g) + Cl^-(s) \rightarrow ClNO_2(g) + NO_3^-(s) \qquad (11)$$

This reaction produces a photolabile species, nitryl chloride ($ClNO_2$), providing a connection between nitrogen oxide pollution and halogen activation, with the yield of $ClNO_2$ approaching 100% (*37*). Ravishankara and co-workers (*57, 58*) showed that this reaction can occur in the atmosphere on aerosol particles of low to moderate chloride content, and $ClNO_2$ may further react with $Cl^-$ to form $Cl_2$. Although additional reaction channels forming ClNO and HOCl may exist (*59*), only high levels of $ClNO_2$ and $Cl_2$ have been observed in the lower stratosphere and troposphere, and $ClNO_2$ is the dominant active chlorine species (*38, 39, 60, 61*). Thus, we will only consider the reaction (11) as the predominant mechanism for converting $Cl^-$ ions trapped on atmospheric particles into active chlorine species in the gas phase. The reaction (11) is important in the chemistry of the upper troposphere and lower stratosphere due to its rapid liberation of highly reactive Cl free radicals upon photolysis. In addition, of particular interest is the reaction (11) in controlling the lifetime of adsorbed $Cl^-$ on the surfaces and hence surface charging of cloud or aerosol particles, as will be explored later.

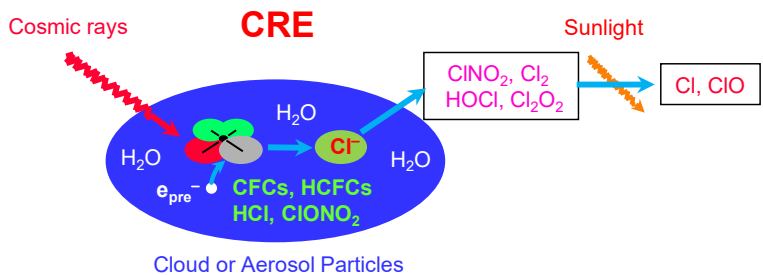

**Fig. 1.** Major reactions in the cosmic-ray-driven electron-induced reaction (CRE) mechanism of ozone depletion. A cosmic-ray driven charge-induced adsorption of a halogen-containing molecule (CFCs, HCFCs, HCl, and $ClONO_2$) onto the charged surface of solid or liquid cloud or aerosol particles and the subsequent dissociative electron transfer (DET) reaction leads to the formation of mainly a $Cl^-$ ion that is trapped at the surface. The $Cl^-$ ion is then converted by reaction with reactive nitrogen species (mainly $N_2O_5$) to release photoactive $ClNO_2$ ($Cl_2$ or HOCl or ClNO) into the gas phase. The photolysis of the latter rapidly results in a Cl free radical to cause ozone destruction via the well-known Cl-catalyzed reaction cycles. Modified from Lu (*17, 62, 63*).

As reviewed previously (*16, 17*), laboratory measurements and field observations have provided a sound physics and chemistry foundation for the proposed CRE mechanism for the formation of an ozone hole, which is schematically shown in Fig. 1. Furthermore, the fingerprints of the CRE mechanism have also been found: the 11-year cyclic variations of ozone loss and associated cooling in the Antarctic are only observed in the lower stratospheric region where the CR ionization rate (the CRE reaction) is large, decrease with increasing or decreasing altitudes and disappear in the upper stratosphere (*63*). It must be noted that the observed 11-year cyclic variations disagree with the photochemical models, which predict only small 11-year cyclic variations of solar-modulated ozone *production* in the *upper* stratosphere involving no (photo)chemistry of ODSs (*64*). In fact, the CRE-predicted existence of 11-year cyclic variations of ozone loss and stratospheric cooling was not unchallenged by the atmospheric chemistry community until very recently (see the author's recent paper (*4*), and references therein).

**Towards Quantitative Understanding of Global $O_3$ Depletion by the CRE Mechanism**

To achieve a quantitative understanding of vertical profiles of $O_3$ depletion, we will need to obtain quantitative estimates of the following three major steps in atmospheric processes: electrons produced at surfaces of ice or aqueous cloud or aerosol particles under CR ionization, charged-induced adsorption of ODSs on particle surfaces, and the yield of gas-phase halogen (Cl) atoms converted from halogen anions ($Cl^-$) produced by DET reactions of adsorbed ODSs on the surfaces to cause $O_3$ loss through the halogen-catalyzed reaction cycles.

*Atmospheric distribution of prehydrated electrons ($e_{pre}^-$)*. First, the flux of $e_{pre}^-$ at the surface of cloud or aerosol particles, $\Phi_e$, can be calculated from the measured CR fluxes ($\Phi_{CR}$) and CR energetics in the stratosphere and troposphere, which are available (*65*). It should be reasonable to assume that all cloud or aerosol particles have a major composition of $H_2O$. In radiation chemistry of water or $H_2O$ ice, the G value (the number of generated/altered species per 100 eV ionizing radiation energy deposited) for the production of $e_{pre}^-$ is known to be ~4.8 (*66, 67*). Thus, we obtain

$$\Phi_e = 4.8\Phi_{CR}(\bar{E}_{CR}/100) \qquad (12)$$

where $\bar{E}$ is the most effective energy of ionizing CRs in the atmosphere. Note that despite the primary galactic CRs (GCRs) having very high energies up to several tens of GeV, the atmospheric ionization at altitudes below 50 km arises mainly not from the primary GCR particles but from secondaries of a nucleonic–electromagnetic cascade initiated by primary GCRs in the atmosphere (*65, 68*). For a certain latitude, the maximum ionization produced by CR particles in the atmosphere depends on the altitude and phase of the solar cycle. This $\bar{E}$ peak moves towards higher energies with decreasing altitudes, and moves slightly to higher energies with increasing solar activity or decreasing latitudes due to the hardening of primary GCR spectrum. The $\bar{E}$ value is about 1.0-2.2 GeV from solar minimum to solar maximum in the polar stratosphere,, increasing to about 3 GeV at ~3 km altitude (*68*). In the equatorial stratosphere, $\bar{E}$ should be slightly larger ($\geq 2.2$ GeV) and the solar modulation should be weakest due to the highest energies of primary GCRs reaching equatorial regions (the geomagnetic effect).

Using the measured CR fluxes varying with altitude (*65*) and approximately $\bar{E}$=1.0, 2.2 and 1.3 GeV for polar, tropical and mid-latitude stratospheres respectively, we obtain the altitude profiles of $e_{pre}^-$ that would be produced at the surfaces of ice or aqueous particles under CR ionization, as shown in Fig. 2a. These vertical profiles, however, must be corrected by the density distributions of cloud or aerosol particles if one wants to get the actual distributions of the $e_{pre}^-$ production rate in the stratosphere and troposphere.

Since the discovery of the Antarctic ozone hole, stratospheric clouds or aerosols have been intensively studied (*69-71*). Particularly Adrian et al. (*69*) made a delicate study, measuring and characterizing PSCs and aerosols in the winter lower stratosphere at McMurdo Station, Antarctica (78°S). The stratospheric particles which were observed above 11 km at temperatures below 198 K and were divided into four classes based on their scattering properties and particle sizes. Namely, volcanic aerosol and nondepolarizing hydrated nitric acid particles were predominant with surface area densities in the range of 10-30 and 20-50 $\mu m^2$ $cm^{-3}$ respectively, in the polar stratosphere below 16-17 km; above this altitude, depolarizing hydrated nitric acid and ice clouds were predominant with surface area densities of 5-15 and 5-100 $\mu m^2$ $cm^{-3}$. It is known that the Arctic polar vortex in winter is much less stable in terms of PSC formation than its Antarctic counterpart and the year-to-year variability in the PSC spatial volume in the Arctic is much larger. In the Antarctic, the PSC season is longer and more regular with presence of PSCs every year from mid-May to early October, while in the Arctic, PSCs may form from December to March but may not occur in any of these months. On average, there were about 14 times more PSC occurrence during a season in the Antarctic than in the Arctic (*71*). In terms of PSC compositions, the season-long mean vertical profile of relative composite spatial coverage (composition-specific area normalized by total PSC area) is also very different between the two Poles (*71*). In the Antarctic, solid nitric acid trihydrate (NAT) particles are the predominant PSC composition (taking nearly 60%) with the peak around 22 km and there are significant percentages (nearly 25%) of $H_2O$ ice particles below 24 km. In contrast, NAT mixtures (taking nearly 80%) peak at around 16 km and drop to about 20% in the upper troposphere near 12 km, whereas ice (cirrus) particles have the maximum (nearly 80%) near 12 km and drop to very low fractions (less than 5%) above 16 km in the Arctic (*71*).

The formation of atmospheric particles such as PSCs and aerosols should depend on temperature and pressure. To simulate the surface area densities of cloud or aerosol particles in the global stratospheric and troposphere, we simply adopt an empirical equation

$$\mu = \mu_0(P/P_0)\exp[-38(T-T_0)/T] \qquad (13)$$

where $\mu_0$=50 $\mu m^2$ $cm^{-3}$, $P_0$=43.72 mb and $T_0$=192 K. Our simulated results of particle surface area densities by Eq. 13 are shown in Fig. 2b. Interestingly, the surface area densities are calculated to be 20-56 $\mu m^2$ $cm^{-3}$ at 10-25 km for cloud or aerosol particles in the winter Antarctic stratosphere, which are in good agreement with the measured ranges (*69, 70*). Our simulated results of atmospheric particles in the tropics give the maximum surface area density of 37-45 $\mu m^2$ $cm^{-3}$ at 16-18 km, rapidly decreasing to less than 10 $\mu m^2$ $cm^{-3}$ at higher or lower altitudes due to the sharp rises in temperature. This is reasonable, given that the measured temperature in the tropical lower stratosphere at 16-18 km is as low as 196-197 K, similar to the winter Antarctic lower stratospheric temperatures (*1, 72*). The calculated $\mu$ values are 1-3 $\mu m^2$ $cm^{-3}$ at midlatitudes and 1-5 $\mu m^2$ $cm^{-3}$ in the Arctic (approximately 10 times lower than in the Antarctic) due to much higher stratospheric temperatures. Our simulated results in Fig. 2b show overall agreement with the above-described observations, particularly for the differences between the Antarctic and the Arctic.

The altitude distributions of the $e_{pre}^-$ production rates ($=\mu\Phi_e$) in the stratosphere and troposphere are shown in Fig. 2c. Of particular interest is that the simulated particle surface area density in the winter Arctic peaks at the attitude around 10 km, arising from the combined effects of temperature and air density/pressure (Fig. 2b), whereas the corrected $e_{pre}^-$ production rates in all regions, including the Arctic, peak at 16-20 km (Fig. 2c).

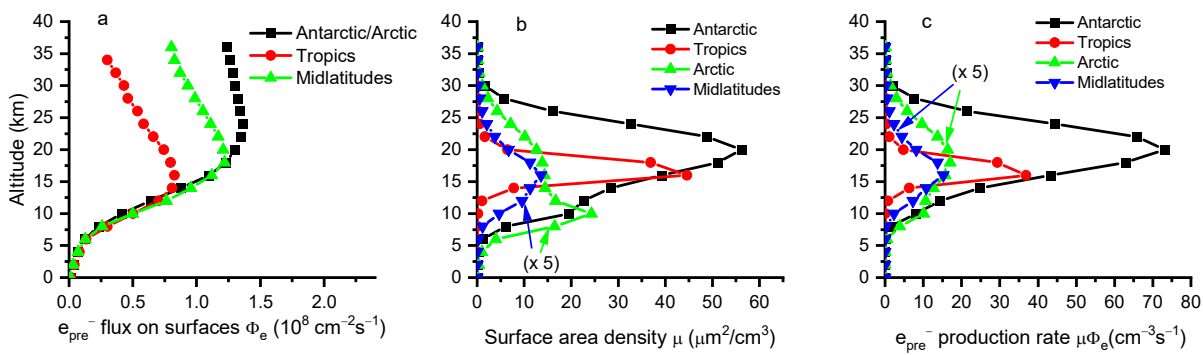

**Fig. 2.** Calculated altitude profiles of (a) the prehydrated electron flux on surfaces of cloud or aerosol particles, (b) the surface area density of the particles, and (c) the prehydrated electron production rate ($\mu\Phi_e$) in the atmosphere, in the winter Antarctic/Arctic and the annual tropics and midlatitudes.

*Charge-induced adsorption of ODSs on atmospheric particle surfaces.* To find the amounts of charge-induced adsorption of ODSs such as CFCs and HCFCs that are non-sticky to an uncharged ice surface, we need to determine the surface charge density on the surface of atmospheric particles. A critical step is to estimate the lifetime of charges on the particle surface. Although the initial process of the CR ionization is the instantaneous production of electrons, electrons are rapidly captured by molecules via DEA or DET to yield anions. As mentioned above, the DET reactions of halogenated ODSs with $e_{pre}^-$ are extremely effective to produce $Cl^-$ ($Br^-$ or $I^-$) ions, most of which are trapped at the surface by the image potential. In addition, electron attachment to other atmospheric species can also lead to the formation of molecular anions such as $N_2^-$ and $O_2^-$, which can effectively trap charges on the particle surface, as we observed on ice surfaces (*46*). Indeed, van Driel and co-workers (*9, 10*) observed that ambient oxygen serves as an effective trapping catalyst in enhancing electron trapping in ultrathin $SiO_2$ films, and that surface charging can then induce effective adsorption of non-sticky and non-polar gases (such as He, Ar, $H_2$, $O_2$, $N_2$, CO) onto the surface at room temperature. The results are well explained by their proposed electrostatic model based on coupling between a negative charge at the surface and the electric-field induced dipole of the adspecies. A similar process is reasonably expected to occur on surfaces of cloud or aerosol particles under CR ionization, on which $Cl^-(Br^-/I^-)$, $N_2^-$ and $O_2^-$ can serve as trapping catalysts. Thus, the surface charge density $\sigma_e$ can simply be written as

$$\sigma_e = \Phi_e \tau_{ion} \qquad (14)$$

where $\tau_{ion}$ is the lifetime of trapped anions at the surface of cloud or aerosol particles. Note that Eq. 14 takes no account of the cumulative effect of anions on the surface, which will lead to a much higher surface charge density.

Now, with the surface charge density given by Eq. 14, we can derive the amounts of ODSs adsorbed on atmospheric particle surfaces. For gas adsorption on the electron-trapped $SiO_2$ surface, van Driel and co-workers (*10*) suggested a universal bonding mechanism—an electrostatic bonding in which the adsorbed species become polarized in the electric field established by the negative charges on the surface and the interaction energy depends on the polarizability of the adparticle. This mechanism well explained their experimental results with the surface irradiated by a femtosecond laser, which was estimated to yield an electron flux at the Si-$SiO_2$ interface of $10^{13}$–$10^{15}$ cm$^{-2}$ s$^{-1}$ (*9*). A similar magnitude of electron flux generated by alkali metal (potassium) pre-deposition on ice was also estimated in our experiment, which showed strong charge-induced adsorption of non-polar $CCl_4$ on $H_2O$ ice surface (*17*). As this electron flux are expected to much higher than the ones produced by CR fluxes at the surface of atmospheric particles expressed in Eq. 14, the electric-field induced polarizability of the adspecies is unlikely to be significant in the stratosphere and troposphere. Thus, we will only consider the electrostatic bonding between the negative charges on the surface and the intrinsic dipole of the adspecies (an ODS), noticing that most of the ODSs, except $CCl_4$, are polar molecules with an electric dipole moment. For interaction of the adspecies with a single-charged surface anion of charge *e,* the binding (interaction) energy simply becomes

$$E_b = \delta e / 4\pi r_0^2 \qquad (15)$$

where e is the electron elementary charge, $\delta$ is the electric dipole moment of the adspecies, and $r_0$ is the equilibrium adsorption distance. This distance is approximately the sum of the radius of the surface anion and the van der Waals radius of the adspecies. More precisely, the distance $r_0$ should be determined by quantum-chemistry calculations with the attractive interaction, electron-electron repulsion, and Pauli exclusion taken into account (*10*). The binding energy in Eq. 15 should be doubled for some ODSs such as $CF_2Cl_2$ and $CFCl_3$, which are expected to have two Cl atoms directed toward the surface and experience the electrostatic field caused by two surface charges. Given the measured dipole moments of ODSs (CFCs, HCFCs, HCl, $ClONO_2$, and $N_2O_5$) and taking the surface charge species to be $Cl^-$, we calculate by Eq. 15 that the binding energies are 270 meV and 464 meV for $ClONO_2$ and $N_2O_5$ respectively, and are in the range of 342-540 meV for CFCs, HCFCs and HCl, from CFC-11 with a smallest $\delta=0.46$ Debye (D) to HCFC-142b with a largest $\delta=2.14$ D. The actual $E_b$ value will be much larger if one considers interactions with all the other charges on the surface and if one also includes field-induced chemisorption effects (*10*). For simplicity, we approximately use the binding energy $E_b^i \approx 400$ meV for all HCl, CFCs and HCFCs and the above-given binding energies for $ClONO_2$ and $N_2O_5$ in our calculations of the coverages of these ODSs on the surface of atmospheric particles.

Note that charge-induced adsorption based on the charge-dipole coupling, which can considered as *targeted reactive adsorption*, is very different from conventional adsorption/uptake of gases on the particle surface according to Henry's law. The latter was proposed to be the rate-limiting step for the heterogeneous chemical reaction of HCl and $ClONO_2$ on the surface of PSC ice (*27-29*). For the latter reaction, the reactive uptake coefficient (γ) or reaction probability is very complicated to determine, depending in complex steps on the gas, interface, and condensed phase environment of the particles and requiring the development and application of complicated models with multiple parameterizations to interpret γ (see a recent review by Wilson et al. (*73*), and references therein). From Langmuir's adsorption equation, in contrast, the charge-induced adsorbed surface density $[ODS]_{ad}^i$ of an ODS species with a partial pressure $p_i$ in the atmosphere onto the particle surface is simply given by

$$[ODS]_{ad}^i = \sigma_{max}\theta_{ODS}^i = \sigma_e\theta_{ODS}^i \qquad (16)$$

Here $\sigma_{max}$ is the maximum surface concentration of the ODS (in molec/cm$^2$), which is simply the surface charge density $\sigma_e$ in this charge-attracted adsorption, and $\theta_{ODS}^i$ is the coverage of the ODS species on the surface, defined as

$$\theta_{ODS}^i \equiv K_{eq}^i p_i/(1+K_{eq}^i p_i) \qquad (17)$$

where $K_{eq}^i$ is the equilibrium constant (in liter/mole) governed by two opposing adsorption-desorption steps and is equal to the ratio of their rate constants

$$K_{eq}^i \equiv k_{ad}^i/k_{de}^i = \exp(E_b^i/k_BT). \qquad (18)$$

*The yield of halogen (Cl) atoms converted from halogen anions (Cl$^-$) produced by DET reactions of ODSs adsorbed on particle surfaces.* To find the Cl radical yield, we need to know the lifetimes of $e_{pre}^-$ and anions on the particle surface in the atmosphere, $\tau_e$ and $\tau_{ion}$. Long-lived trapped electrons on ultrathin ice films were implicitly observed in our electron-trapping experiments (*13, 15, 50, 74*). Subsequently, Baletto et al. (*75*) found the quite interesting result by their first-principles molecular dynamics simulations that very stable surface-bound states for trapped electron can exist on ice surface at temperatures of 150-200 K, similar to the temperatures measured in the lower stratosphere in the winter Antarctic/Arctic or the all-season tropics. Baletto et al. also found that the structural rearrangement induced by an excess electron at the ice surface traps the electron in a very stable surface state and hinders its decay into the bulk through the buildup of a subsurface electrostatic barrier. Moreover, Bovensiepen et al.(*76*) made a remarkable experiment by femtosecond time-resolved two-photon photoelectron spectroscopy, providing direct observations of very long-lived trapped electrons with a lifetime up to minutes (10$^3$ s) at the crystalline ice surface. In their time- and energy-resolved experiment, notably the stabilization of the trapped electrons was monitored continuously in timescales over 17 orders of magnitude from femtoseconds to minutes, exhibiting almost no shift in energy and no decay in yield of the trapped electrons up to 10$^{-2}$ s. Bovensiepen et al. also performed first-principle calculations, showing that the electrons are stabilized at pre-existing structural traps on the surface of the crystalline ice with the electron wave functions being effectively screened from the underlying bulk, leading to the observed extremely long lifetimes. For simplicity in our current calculations of global ozone loss, we will use $\tau_e \approx 1.0 \times 10^{-2}$ s. It is obviously expected that larger values of ozone loss can be obtained if one uses a longer lifetime $\tau_e$.

The lifetime of anions is related to the loss process of anions in the atmosphere. In the general atmosphere below 50 km, the major sink of negative ions is recombination with positive ions and the rate constant for the recombination is essentially independent of species (*52, 53*). In the upper stratosphere, the recombination lifetime of the ions is very long, about $10^4$ s. With lowering altitudes, the lifetime decreases until about the tropopause and increases again in the troposphere down to the ground, resulting in a minimum in the lowermost stratosphere (*77-80*). The ion-ion recombination lifetime in the stratosphere is approximately $10^2$–$10^4$ s, estimated from the measured recombination coefficients (*78, 80*). On the particle surface, we expect that the ion-ion recombination is dominated over by the reaction of anions with atmospheric species (mainly reactive nitrogen species, NOx), as expressed in reaction (11). Slightly different from the previous proposal of the surface ionic reaction (*38, 39, 54-58, 60, 61*), here we propose that $N_2O_5$ being a polar molecule is first adsorbed onto the charged particle surface via the targeted electrostatic attraction *rather than random collisions*, and then reacts with the already trapped $Cl^-$ to yield a photoactive $ClNO_2$. The second step is similar to a surface reaction scheme recently proposed to describe trace gas uptake and reaction with applications to aerosols and microdroplets by Wilson et al (*73*). If we only consider the conversion reaction with $N_2O_5$ via reaction (11) for all trapped $Cl^-$ ions, the lifetime of trapped anions is governed approximately by the adsorbed $N_2O_5$ concentration on the particle surface, $[N_2O_5]_{ad}$

$$\tau_{ion} = [k_{Cl^-}\mu[N_2O_5]_{ad}]^{-1} \qquad (19)$$

where $k_{Cl^-}$ is the rate constant for the reaction (11)

$$k_{Cl^-} = 1.07\times10^{-15}\exp(-E_a/k_BT) \text{ (cm}^3\text{s}^{-1}\text{)} \qquad (20)$$

with the calculated activation energy being $E_a=-354$ meV and the measured $k_{Cl^-}$ value at 300K being $k_{Cl^-}=9.4\times10^{-10}$ cm$^3$s$^{-1}$ (*81*). Combining Eqs. 14, 16 and 19, we obtain a more meaningful expression of $\tau_{ion}$

$$\tau_{ion} = [k_{Cl^-}\mu\Phi_e\theta(N_2O_5)]^{-1/2} \qquad (21)$$

where $\theta(N_2O_5)$ is the coverage of adsorbed $N_2O_5$ on the particle surface given by Eqs. 17 and 18, depending on the gaseous $N_2O_5$ concentration and the binding energy of $N_2O_5$ with the charged surface. It is seen that in addition to these quantities, several other factors also influence the lifetime of trapped anions at the surface of atmospheric particles. These include the particle surface area density, the CR-produced surface electron flux, and the rate constant for the reaction of $N_2O_5$ with the anion. Like long-lived trapped electrons, anions trapped at the surfaces of atmospheric particles are expected to have a longer lifetime than the ion-ion recombination lifetime in the gas phase or the bulk. On the other hand, it is also expected that owing to involving surface reactions, the lifetime of surface trapped anions will have a much stronger dependence on atmospheric temperature than the free anions.

To find the lifetime of trapped anions on the surface by Eq. 21, an additional important factor is the 'denoxification' reducing the concentrations (partial pressures) of gaseous $NO_2$ and $N_2O_5$ in the coldest stratosphere, which is dependent on the surface area density $\mu$ of the particles

$$[N_2O_5] = [N_2O_5]_0(1-k_{den}\mu) = \alpha^{-1}[N_2O_5]_0 \qquad (22)$$

where $k_{den}/\alpha$ is a reducing constant or factor of denitrification and $[N_2O_5]_0$ is the gas-phase $N_2O_5$ concentration if there are low densities of particles ($k_{den}\mu \ll 1$). High-quality satellite data of $N_2O_5$ data with a good altitude resolution are needed to determine the factors of denitrification. The NASA's HIRLDS (v7) $N_2O_5$ data are unique in having a vertical resolution of ~ 1 km, but they only cover the stratosphere at altitudes above 20 km and latitudes 63°S-80°N for the period January 2005-March 2008 (*82, 83*), with no data available for the main Antarctic (60°S-90°S). To overcome this challenge, we fit Eq. 22, with the particle surface area density $\mu$ given by Eq. 13, to the HIRLDS satellite data in the tropics and obtain the constant $k_{den}=1.5\times10^6$ cm. This $k_{den}$ value gives the factors $\alpha$ of reduction in $N_2O_5$ concentration in the lower stratosphere at 10-25 km to be 1.0-3.0 in the annual tropics, relative to the measured $N_2O_5$ concentrations in March. The extending of this $k_{den}$ value to the winter Antarctic is justified by the observation that the lower-stratospheric temperatures in the tropics are very close to those in the winter Antarctic (*1, 72*). The thus obtained $\alpha$ values lie in 1.3-6.4 in the winter Antarctic. The $\alpha$ values determined from the available HIRLDS satellite data are ~3.0 for the winter Arctic and ~2.0 for annual midlatitudes in the lower stratosphere. It is known that the observed winter NOx is reduced by up to a factor of ~10 with respective to summer observations in the 20-27 km altitude region (*35*). Given the differences in lower-stratospheric temperature in these regions, these denitrification factors are very reasonable and will be used in our calculations of ozone loss rates.

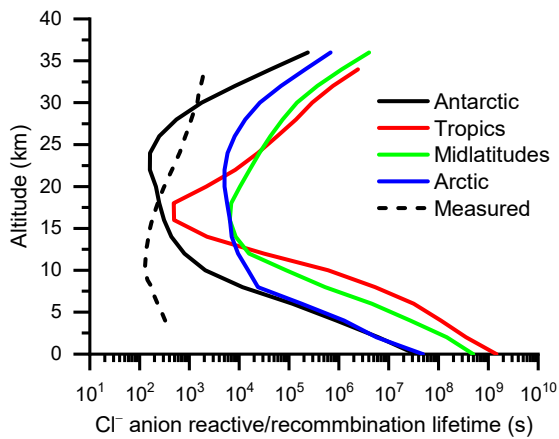

**Fig. 3.** Calculated altitude profiles of the reactive lifetime of Cl⁻ anions trapped at atmospheric particle surfaces in the winter polar regions and the annul tropics and midlatitudes. Also shown is the measured altitude profile (the dash curve) of the ion-ion recommbination lifetime in the general atmosphere.

As plotted in Fig. 3, our calculations of the reactive Cl⁻ anion lifetimes show very interesting results that despite the expected differences from the ion-ion recombination lifetimes in the gas phase, the lifetimes of surface trapped Cl⁻ given by Eqs. 20-22 well reproduce not only the measured ion lifetimes of $10^2$–$10^4$ s in the lower stratosphere but also the observed altitude profiles, exhibiting a minimum lifetime in the lower stratosphere (*78-80*). Moreover, it is particularly interesting to note that the use of our simulated factors of denitrification by Eq. 22, together with the particle surface area densities given by Eq. 13, also leads to our calculated ozone depletion rates in excellent agreement with those using the actual satellite-measured $N_2O_5$ concentrations available at altitudes above 20 km except the Antarctic, as will be presented later.

In the CRE mechanism of ozone depletion, the DET reaction (10) is clearly the rate-limiting step, which is solidly indicated from the above-mentioned observations of pronounced 11-year cyclic variations of ozone loss and associated cooling in the lower stratospheric region where the CR ionization mainly occurs, especially in polar regions (*4, 17, 63*). As mentioned above, it is well known that $ClNO_2$ formed from the Cl⁻ conversion reaction (11) is rapidly photolyzed to yield a Cl atom with sunlight (see reaction 3b). As a result, the average diurnal sum concentration of Cl atoms yielded from the CRE mechanism of all ODSs, expressed as a volumetric concentration by the particle surface area density μ in the atmosphere, is given by

$$[Cl] = \sum_i \mu k_e^i \, [ODS]_{ad}^i \Phi_e \tau_e \qquad (23)$$

where $k_e^i$ is the DET cross section for an adsorbed ODS species with $e_{pre}^-$ produced by CR ionization on the particle surface. Note that Eq. 23 actually takes into account of the catalytic (cumulative) effect of Cl⁻ on the adsorption of an ODS onto the surface by assuming that the ODS is not depleted during the lifetime of $e_{pre}^-$. Substituting Eqs. 14 and 16 into Eq. 23, we obtain

$$[Cl] = \sum_i \mu k_e^i \, \tau_e \tau_{ion} \theta_{ODS}^i \Phi_e^2 \equiv \sum_i k^i \theta_{ODS}^i \Phi_e^2 \qquad (24)$$

where $k^i \equiv \mu k_e^i \tau_e \tau_{ion}$ is defined as the effective (volumetric) DET coefficient of an ODS (adsorbed on the particle surface) in the atmosphere. This equation, together with Eq. 8, provides a complete quantitative estimate of global ozone depletion in the atmosphere.

Provided with the measured DET cross sections (*11, 13-15, 17*) and altitude profiles (*35, 84*) of ODSs (CFCs, HCFCs, HCl, $ClONO_2$, and $N_2O_5$), the calculated values of $\Phi_e$, μ, $\theta_{ODS}^i$, and $\tau_{ion}$ by Eqs. 12, 13, 17-18 and 20-22 respectively with the binding energies $E_b$ given in the above, and $\tau_e \approx 1.0 \times 10^{-2}$ s, we can now use Eqs. 8 and 24 to calculate the altitude profiles of ozone depletion rate at various regions of the globe. The good reproductions of measured values including surface area densities, magnitudes of denitrification, and anion lifetimes by our simulated results mentioned above give us confidence to make quantitative calculations of global ozone loss.

**Observations and Quantitative Calculations of Global Ozone Depletion**

*Observations.* There are two available satellite datasets covering the periods starting from the 1970 or late 1970s, NASA/NOAA SBUV and NASA's SAGE I/II data sets. However, the SBUV profile data are not useful particularly in the lower stratosphere and troposphere, due to the instrument's limited vertical resolution, which is 6–7 km near 3 hPa and degrades to 15 km in the troposphere. Therefore all satellite merged profile records rely mostly on SAGE I/II data for pre-2000 periods, as reviewed in the recent LOTUS Report (*7*). The NASA's GOZCARDS is a combination of various high-quality space-based monthly zonal mean ozone profile data (*6*). Its merged data set was constructed by combining various satellite ozone measurements after removing systematic biases with respect to the SAGE II data set, which was used as a reference standard (*6*). It is worth noting that among long-term merged satellite data sets (SBUV MOD/COH, GOZCARDS, SWOOSH, SAGE-OSIRIS/CCI/MIPAS-OMPS), GOZCARDS is the only merged satellite dataset that extends to the lowermost stratosphere and covers the early satellite era starting in 1979 (*7*). In GOZCARDS, the recommended data range is 215 hPa to 0.2 hPa outside the tropics; it is 100 hPa to 0.2 hPa at tropical latitudes to ensure only stratospheric data are considered. The main variations of the GOZCARDS merged ozone values before 2005 are governed by the HALOE and SAGE II ozone data sets, which agree quite well with zonal mean differences generally within 5% for 1.5 to 68 hPa at midlatitudes and for 1.5 to 46 hPa in the tropics; relative biases are larger outside those ranges and increase to ~10% near the tropopause (*6*).

Here, measured altitude profiles of ozone depletion in polar regions, the tropics and mid-latitudes from the ground-based ozonesonde (TOST_SM) and satellite (GOZCARDS _source and _merged) datasets are shown in Fig. S1 and Fig. 4, to compare with our theoretically calculated results. Owing to the sparseness in ground-based measurement stations in the 1960s and 1970s, potential errors could arise from the data in the pre-1980 period, compared to the 1980s with an improved data coverage. For this concern, we show the data of ozone loss in the 2000s with respect both to the 1960s in Fig. S1 and to the 1980s in Fig. 4, as we presented in recent papers (*1, 4*). Prior to 1979, there lacked reliable satellite data. Thus, we similarly present the GOZCARDS data of ozone loss in the 2000s relative both to the 3-year mean satellite-measured data for the first three years 1979-1981 as the starting point of the satellite era in Fig. S1, as we and others presented previously (*4, 85*), and to the decadal mean data in the 1980s in Fig. 4. For measured ozone data, the standard deviations were reported to be typically within 10% of the means (*3, 6*). To consider the instrumental errors and other uncertainties such as bias uncertainties and altitude correction uncertainties, we apply a total error bar of ±20% to the mean values. To derive the measured and theoretical vertical profiles of ozone loss rate averaged in a diurnal cycle (per day), we take the lifetime of ozone in the stratosphere and troposphere to be about 30 days (*2, 3, 35*). This leads to the maximum average diurnal ozone depletion rate of about 2.5% per day in the springtime Antarctic ozone hole in September-October-November (SON) (Fig. S1a), which is consistent with the values reported in the literature (*8, 34*), and of 2.3% per day in the annual tropical ozone hole (Fig. 4f). As plotted in Fig. S1 and Fig. 4, the results of vertical profiles of ozone loss from the independent TOST and GOZCARDS datasets show overall consistency. This is very similar to the previous finding by Randel et al. (*5*), though the latter focused only on northern hemisphere midlatitudes. However, some differences between the datasets are also seen, especially for the vertical profiles of ozone loss in the spring Arctic in March-April-May (MAM) and the annual tropics. It is well known that the year-to-year variability in PSC formation and ozone loss in the Arctic is much larger than in the Antarctic, with very large dynamic variabilities for most winter and spring days. This factor could contribute to the large discrepancy between TOST and GOZCARDS data for ozone loss in the Arctic. In the tropics, the TOST data exhibit a very broad peak in altitude profile of ozone loss at altitudes 8-25 km, whereas the GOZCARDS data show a much narrower peak centered around 16 km though no satellite data are available below 14 km. This large discrepancy might be caused by the domain-filling trajectory approach used in generating the TOST dataset. There are overall excellent agreements between GOZCARDS raw (source) and merged datasets, while there are also some minor differences between the two datasets. For example, one can see in Fig. 4a that the ozone loss rate in the Antarctic ozone hole in the merged dataset is lower than that in the raw dataset by about 0.2% per day at ~15 km altitude and even becomes negative at about −0.2% per day, indicating an ozone 'increase', in the middle stratosphere 22-32 km. This artificial result and the other relatively minor differences between the GOZCARDS raw and merged datasets were likely caused by the bias removal and averaging in generating the merged data. Nevertheless, the observed data from both ground-based TOST and satellite-measured GOZCARDS datasets should reasonably represent ozone trends from the 1960s/1980 or the 1980s to the 2000s.

*Theoretical calculations*. For our theoretical study on ozone destruction by the CRE mechanism, we have calculated the average diurnal Cl atom concentration produced by the CRE mechanism via Eq. 24 and then the ozone loss rate via integration of Eq. 8 over a 24-hour period to yield the percentage of $O_3$ depleted in each diurnal cycle (*36*). Here we present our calculated results of average diurnal $O_3$ loss rates for the Antarctic (60°S-90°S) and the Arctic (60°N-90°N) in the spring season (SON for the South Pole and MAM for the North Pole), the annual tropics (30°S-30°N), and the annul midlatitudes (30°S-60°S and 30°N-60°N) for the 2000s with respect to the 1960s/1980 or the 1980s, corresponding to the same periods for observed data from both TOST and GOZCARDS, as shown in Fig. S1 and Fig. 4. As for our theoretical calculations, we have to make the following notes: (i) Since we are interested in studying anthropogenic ozone depletion, the effects of natural halogenated species such as $CH_3Cl$ and $CH_3Br$, which have had only small changes in concentration since 1960s, are not included. Also, $CCl_4$ is not included since it is a non-polar molecule and no electric-field induced polarizability is considered in our current CRE mechanism and therefore the binding energy $E_b$ given by Eq. 15 for this molecule is zero. (ii) We simply assume that the stratospheric concentrations of organic halocarbons (CFCs and HCFCs) and inorganic halogen reservoirs (HCl and $ClONO_2$) above 10 km in the Antarctic/Arctic in the beginning of the winter are approximately 0.75 and 2.5 times the measured concentrations at 30°N in March, respectively, while the corresponding concentrations at mid-latitudes are about 0.85 and 1.5 times; the concentrations in the tropics are approximately the same as those at 30° N. The thus given ODS concentrations are in generally good agreement with the satellite- measured data (*1, 35, 84, 86*). For example, the given HCl concentrations at 22-24 km in the early winter Antarctic/Arctic is 2.02-2.35 ppbv, while the satellite measured value is about 2.0 ppbv at 31.6 hPa (~24 km) at 80°S/N (*86*); the satellite-measured $CF_2Cl_2$ concentrations in the fall Antarctic/Arctic at 16-24 km are about 0.74-0.76 times those in the annual tropics (*1*). The pre-denitrified stratospheric $NO_2/N_2O_5$ concentrations in the Antarctic/Arctic, mid-latitudes and the tropics are respectively about 2.0, 1.5 and 0.83 times the concentrations at 30° N in March (*35, 87*). The above-given denitrification factors in the lower stratosphere at altitudes above 10 km in various regions are used in our calculations. There is no denitrification to occur in the warmer troposphere, in which the gaseous $N_2O_5$ concentration is essentially constant with altitude and is simply fixed at 20, 17.5 and 15 pptv in the Antarctic/Arctic, mid-latitudes and the tropics respectively. Except for the Antarctic, the satellite-measured data of $N_2O_5$ concentrations available at altitudes above 20 km obtained from the NASA's HIRLDS (v7) data set (*82, 83*) are also directly used in our calculations, in addition to those with respect to the measured concentrations at 30° N in March and applying the factors of denitrification. (iii) We assume that there is a lag time of about 10 years for transport and mixing associated with transport from the measured lower tropospheric global mean abundances to the global stratospheric levels of ODSs (*1, 63, 72*). (iv) Since air downward movement occurs in the winter polar vortex, the calculated reactive Cl yield for springtime ozone depletion resulting from the DET reactions on the surfaces of PSC ice particles in the winter Antarctic stratosphere are shifted downwards by 3.2 km, according to the field measurements of PSCs, which showed an altitude displacement of approximately 1.0 km per month (*69*). In contrast, our simulated results by Eq. 13 show that the particle surface area density in the Arctic peaks near the tropopause (see Fig. 2b), and both NAT and ice (cirrus) particles predominantly form in the Arctic lowermost stratosphere at altitudes 12-16 km (*71*), as mentioned above. It is also known that the more irregular underlying surface topography in the Northern Hemisphere leads to stronger upward-propagating wave activity than in the Southern Hemisphere, causing a weaker and more disrupted Arctic vortex compared to the Antarctic (*71*). These factors indicate that unlike the Antarctic, no displacements will be needed to apply to the CRE-produced Cl atom yield in the Arctic. (v) We use the available satellite-measured altitude profiles of stratospheric and tropospheric temperatures in the *winter* Antarctic/Arctic and the annual tropics and mid-latitudes in the 2000s and assume no changes in the temperature profiles from the 1960s or 1980s to the 2000s. In other words, *the only variable is the levels of ODSs (CFCs, HCFCs, HCl and $ClONO_2$) in our time-series calculations*. This is certainly a rough simplifying assumption since ozone depletion is well-known to cause a cooling trend in lower stratospheric temperature, and increased levels of greenhouse gases can also cause a direct radiative cooling in the upper stratosphere particularly at altitudes above 25 km, accompanying a warming in the lower troposphere (*72, 88, 89*). Fortunately, however, unlike the spring season, ozone depletion in the *winter* polar lower stratosphere has been limited and there have been no significant changes in the measured season-mean lower-stratospheric temperatures over Antarctica in the winter (from May to August) or the fall (from March to May) since the 1960s up to date (*4, 16, 17*). Thus, this assumption should have a negligible effect on our calculations of polar ozone depletion. In the tropics and mid-latitudes, however, this assumption may somewhat overestimate lower-stratospheric ozone loss in the 1960s or 1980s and correspondingly underestimate relative ozone loss in the 2000s, due to the relatively low stratospheric temperatures and high tropospheric temperatures in the 2000s being used in all the calculations. Oppositely, the latter may lead to overestimates of ozone loss in the troposphere in the 2000s relative to the 1960s or 1980s. (vi) Finally no enhanced ozone *production* by halogen radicals is included in our CRE calculations. This likely results in overestimates of tropospheric ozone depletion caused by halogenated ODSs in the polluted regions, e.g., at northern midlatitudes. Thus, our calculated ozone loss rates in the stratospheres of the tropics and midlatitudes are the lower limits and in the troposphere are the upper limits.

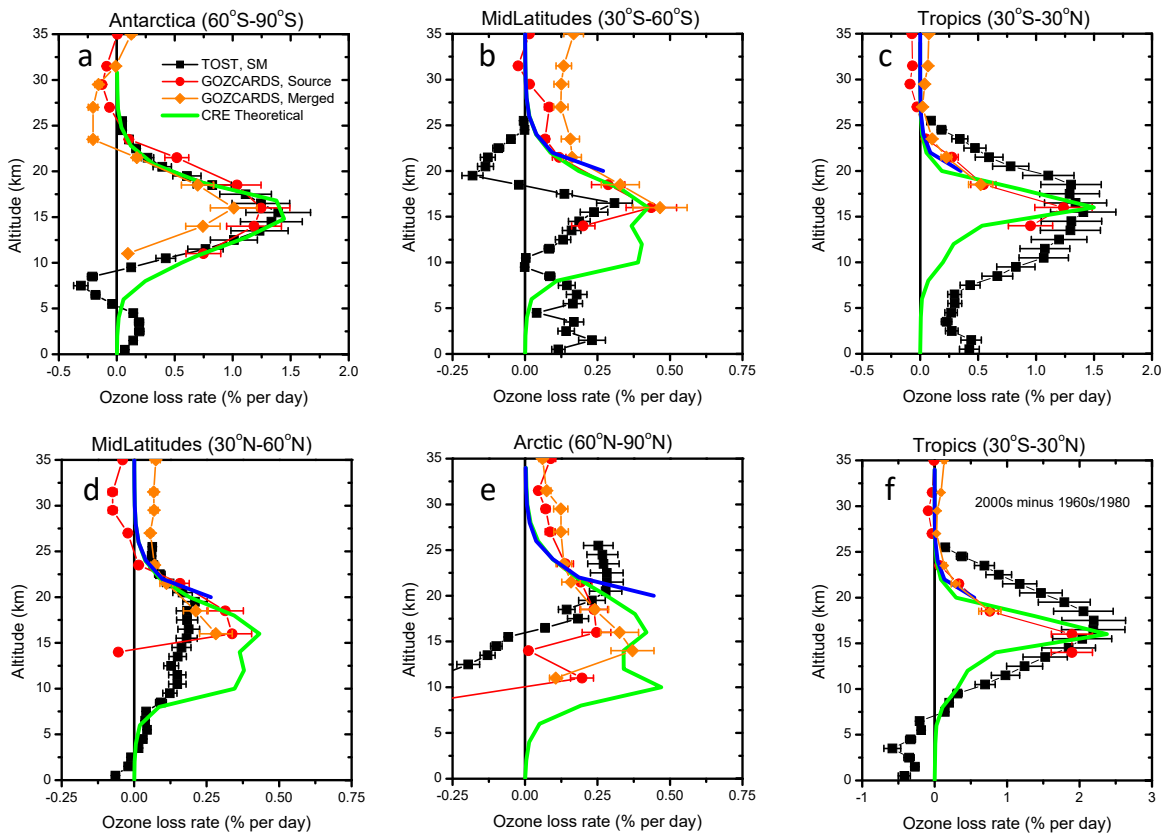

**Fig. 4.** Observed (TOST_SM and GOZCARDS_Source and _Merged) (symbols) and theoretical (green curves) altitude profiles of ozone loss rates in the Antarctic (60°S-90°S), the Arctic (60°N-90°N), the tropics (30°S-30°N), and southern and northern midlatitudes (30°S-60°S and 30°N-60°N), in the 2000s relative to the 1980s (*a-e*). Except for the Antarctic (*a*) in which no HIRDLS data of $N_2O_5$ are available, the theoretical ozone loss rates (the blue curves) calculated directly with the satellite-measured HIRDLS $N_2O_5$ concentrations at altitudes ≥20 km are also shown. In *f*, the results for the tropics are for the 2000s relative to the 1960s (theoretical and TOST) and to the 1980 (GOZCARDS), same as in Fig. S1. Horizontal error bars in observed data are ±20%, approximately 2 times the standard deviations (uncertainties), in the decadal mean values.

With the above notes, we can now discuss our calculated results in Fig. S1 and Fig. 4, which exhibit overall good quantitative agreement with the observed data, particularly with the GOZCARDS satellite data sets. Here we highlight a few key features. (i) The calculated results show overall agreements with observed data in the global stratosphere, exhibiting the better agreement with GOZCARDS data than the TOST ozonesonde data. This is most noticeable for ozone loss in the regions except the Antarctic. (ii) For the ozone depletion rate in the Antarctic stratosphere, nearly equally excellent agreements among calculated results, TOST_SM, and GOZCARDS source data are found. (iii) In the stratospheres of the Arctic and both southern and northern midlatitudes, excellent agreements between calculated and measured results, especially

GOZCARDS data, are found, provided with the well-known large year-to-year dynamic variability in Arctic ozone loss. (iv) In the tropics, remarkable agreement between calculated results and GOZCARDS data is found, both showing a much narrower peak in altitude profile of significant ozone loss rates at 13-20 km than the TOST data. (v) It is also very noteworthy that our calculated ozone depletion rates using the given factors of denitrification show excellent agreement with those calculated directly with the actual satellite-measured data of $N_2O_5$ concentrations available at altitudes above 20 km (Fig. 4b-f), except that the satellite data points at the single altitude 20 km altitude apparently give slightly larger calculated ozone loss rates in the Arctic and at midlatitudes. Moreover, it is worthwhile noting that our CRE theoretical results show nearly equally excellent agreements with observed data in the tropics and the Antarctic/Arctic/midlatitudes. This fact indicates the reliabilities of the observed data and our calculated results on ozone loss in the tropics, which has the largest uncertainties in previous studies (*5, 7, 8*). (vi) In the troposphere of northern midlatitudes and the Arctic, large differences between calculated results and TOST data (no GOZCARDS data available) are seen and the calculated results indeed overestimate tropospheric ozone depletion, as we expected above. (vii) Last not least, our calculated results show very small differences in stratospheric ozone depletion rate between (annual mean) northern midlatitudes and the (springtime) Arctic. Given the known strong seasonal variation in trends over northern midlatitudes in the altitude range 10–18 km, with the largest ozone loss during winter and spring (*5*), the $O_3$ loss rates at northern midlatitudes in winter and spring are likely slightly larger than in the Arctic. This behavior appears surprising as we have been impressed with observations of large Arctic $O_3$ loss in some (not every) springs in the past two decades. However, this result is very unlikely to be artificial, as it is found with great consistency between calculated results and GOZCARDS data, and given an $O_3$ lifetime of ~30 days in the stratosphere, our calculated maximum $O_3$ loss rates of 0.42-0.66% per day at 16 km altitude at northern midlatitudes, which has been most studied, in the 2000s relative to 1960s/1980 or to the 1980s (Figs. S1c and 4d), are in good agreement with the widely accepted values of 7.0-7.3% per decade in the 1980s and 1990s (*5, 8*). For the cause of this surprising result, one can see that the $e_{pre}^-$ production rate at midlatitudes is slightly smaller than that in the Arctic (Fig. 2c) and the lifetime $\tau_{ion}$ of trapped anions on particle surfaces at midlatitudes is nearly identical to that in the Arctic at altitudes around 15 km (Fig. 3), whereas the concentrations of ODSs in the lower stratosphere at midlatitudes are known to be larger than those in the Arctic. The combination of these factors is likely to lead to the small differences in $O_3$ loss rate between the two regions shown in Figs. S1 and 4. Future tests of this result will be of interest, though $O_3$ loss in the Arctic is known to have extremely large interannual variability.

The good agreement between the observed and calculated results in Fig. S1 and Fig. 4 is the most compelling message delivered by this study. Of special note is the good agreement between calculated results and GOZCARDS data in the tropics. Both consistently show a sharp altitude band at 13-20 km of large ozone depletion. These results strongly support our recent discovery of the large and deep tropical ozone hole based on the zonal mean TOST ozonesonde data at the altitude band 13.5-20.5 (14-21) km. In spite of showing the much broader peak in altitude profile of tropical ozone loss at 8-25 km, the TOST data that were used to reveal the tropical ozone hole in our recent study (*1, 4*) lie in the altitude range of 13.5-20.5 km (14-21 km), which well matches the narrow band at 13-20 km of large ozone loss rates in both calculated and GOZCARDS data, as shown in Figs. 4c and 4f. These results not only strongly validate our previous application of the TOST data set in discovering the tropical ozone hole but are strong evidence of the CRE mechanism responsible for global ozone depletion.

Another important finding from our results is the new identified effect of denitrification. It has been known for some time that denitrification is a requirement for observing a large Antarctic ozone hole, which was suggested to reduce the deactivation and removal of ClO radicals by $NO_x$ from the catalytic reaction cycles (*30, 31*). In this study, we have provided a new alternative explanation, that is, denitrification directly increases the lifetime of trapped anions on the surface of atmospheric particles and hence the surface charge density. This can then lead to the enhanced adsorption of ODSs on particle surfaces and the final yield of Cl atoms from the DET reactions of adsorbed ODSs. Importantly, we have presented this new explanation *quantitatively,* straightforward through Eqs. 19, 22 and 24. This new identified effect may play a key role in controlling ozone depletion, as both denitrification and DET reactions simultaneously occur in the winter polar lower stratosphere mainly. In contrast, ClO radicals are predominantly formed in the springtime polar lower stratosphere with the presence of sunlight, during which the level of $ClONO_2$ is actually observed to be at its maximum during the whole year (*17, 35*).

We are now left with theoretical uncertainties arising mainly from three categories: temperature-dependent reaction rate constants ($k_e$ and $k_{Cl^-}$), binding energies between ODSs and the charged surface of atmospheric particles, and lifetimes of $e_{pre}^-$ and anions on the particle surface ($\tau_e$ and $\tau_{ion}$). These lead to an estimated uncertainty of approximately ±20% of the mean values in the determination of $O_3$ loss. This uncertainty is considerably larger than what would be expected from tropical ozone loss due to dynamic changes in BDC, which have been simulated to be 2-3% per decade by CCMs (*90*), corresponding to an ozone loss rate of 0.13-0.20% per day in the tropics in the results shown in Figs. 4c and 4f. In our analyses for lower-stratospheric altitudes at 15–17 km where ozone loss has the maximum, we find that over 80% (88%) of the observed springtime $O_3$ loss rate in the Antarctic (Arctic) and over 96% of the observed annual mean $O_3$ loss rate in the tropics result from the DET reactions of adsorbed CFCs and HCFCs alone, and that the sum of the DET reactions of all adsorbed ODSs including also inorganic chlorine reservoirs (HCl and $ClONO_2$) on surfaces of cloud or aerosol particles leads to catalytic loss rates in good agreement with the observed results in all the polar regions, the tropics, and midlatitudes. The relative contributions from HCl and $ClONO_2$ (mainly HCl) increase with rising altitudes up to about 94% in the middle stratosphere at 33 km in the Antarctic/Arctic and up to 75% at 34 km in the tropics. Thus, our results lead to an important conclusion that both springtime polar and all-season tropical ozone holes observed in the lower stratosphere are predominantly caused by DET reactions of CFCs and HCFCs adsorbed on atmospheric particle surfaces. This directly answers the key question why the large ozone hole with a similar depth to the Antarctic ozone hole is formed in the tropical lower stratosphere where the concentrations of HCl and $ClONO_2$ are very low.

**Maps of Global Ozone Depletion**

To further display the tropical ozone hole, we show global maps of ozone depletion at altitudes 13.5-20.5 km in both the SON season and the annual in the 2000s with respect to both the pre-1980 period (the mean of data in the 1960s and 1970s) and the 1980s, using the TOST_SM dataset with global 3D (i.e. latitude, longitude, altitude) climatology of tropospheric and stratospheric ozone (*3*), in Fig. 5. The maps with the two references similarly exhibit ozone loss in the whole tropical latitude band of 30°S-30°N, fairly uniform in longitude. The pre-1980 reference gives significantly large ozone loss (up to 80%), compared with the 1980s reference. This is expected, as there was significant ozone loss in the 1980s. As expected, significant Antarctic ozone loss, which is also more or less uniform in longitude, is only observed in the SON season (left panels in Fig. 5). Despite the found differences in vertical profile of ozone loss between TOST and GOZCARDS datasets shown in Figs. S1 and 4, the ozone maps in Fig. 5 clearly reveal the spatial scope of the all-season tropical ozone hole at nearly its peak altitude band of 13-20 km, together with the springtime Antarctic ozone hole.

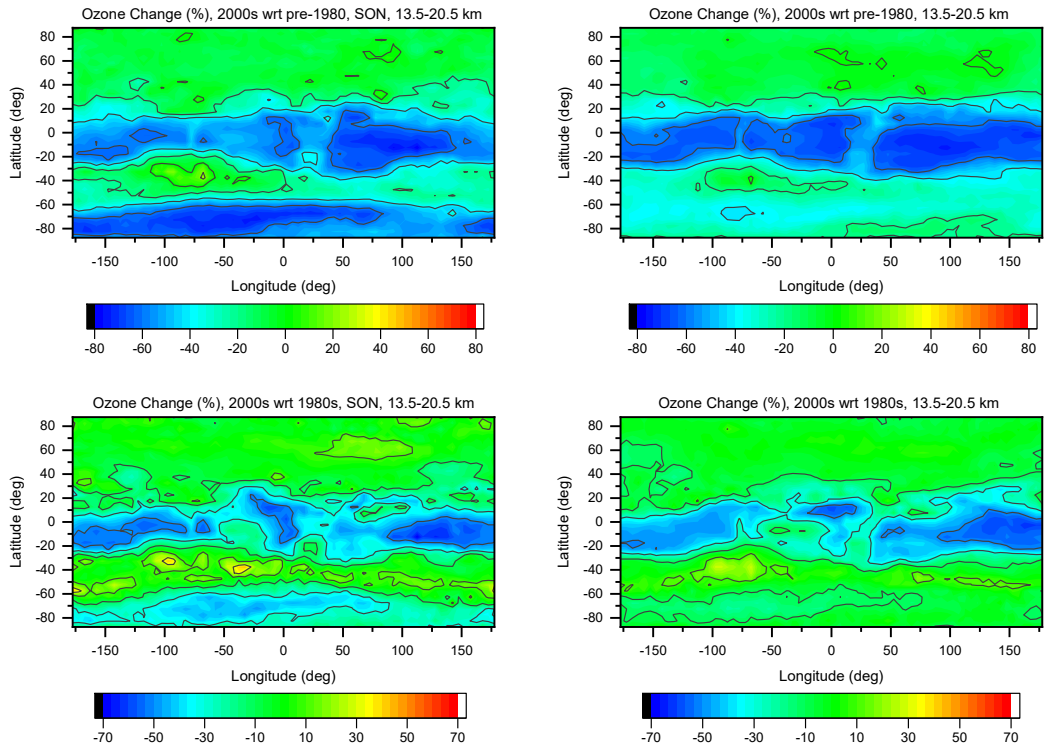

**Fig. 5.** Global maps of observed ozone changes in percentage at the altitude band 13.5-20.5 km in the 2000s relative to the pre-1980 and to the 1980s in both the September-October-November (SON) mean (Left panels) and the annual mean (Right panels), obtained from TOST_SM dataset. The results clearly show the springtime Antarctic ozone hole and the all-season tropical ozone hole with remarkable global symmetries.

**Summary**

A quantitative understanding of global ozone depletion based on the CRE mechanism gives rise to a concise and elegant equation with two major inputs only, the stratospheric concentrations of ODSs and the cosmic-ray flux. Using this equation, our time-series calculations of decadal mean ozone depletion rates in the Antarctic, Arctic, tropics and mid-latitudes with the simplifying assumption taking the variation of halogenated ODSs' concentrations as the sole input impressively yield results in good agreement with observed data, especially with the GOZCARDS satellite data. Excellent agreement among calculated results,

TOST_SM, and GOZCARDS source data is found for the spring Antarctic ozone hole. Remarkable agreement between calculated results and satellite data is found for the all-season tropical ozone hole, which is located within a narrow altitude band at 13-20 km. This agreement puts our recent finding of the largest ozone hole in the tropical lower stratosphere at 14-21 km on a solid ground. Given the known large dynamic variability in Arctic ozone loss, our calculated results also show good agreement with satellite data in the Arctic. Excellent agreement between calculated and measured results, especially GOZCARDS data, is also found in both southern and northern midlatitudes. We have also provided a new insight into the effect of denitrification. Our results show that denitrification can directly increase the lifetime of trapped anions on the surface of atmospheric particles and thus increase the final yield of Cl atoms from the DET reactions of adsorbed ODSs. This new effect is *quantitatively* expressed in the equations, playing a key role in controlling ozone depletion. Interestingly our calculated results, consistent with the GOZCARD data, show very small differences in ozone depletion rate in the lower stratosphere between annual mean northern midlatitudes and the winter Arctic. This behavior is likely due to the combined effect of the prehydrated electron production rate and the ODS concentrations in the stratospheres of the two regions. The maps of global ozone depletion observed from the TOST data set clearly reveal the scope of the largest tropical ozone hole that exists with a remarkable longitude symmetry across the globe since the 1980s.

  We should note that the analytical equation (24) derived in this study seems to generally overestimate ozone depletion in the troposphere, particularly in polluted regions (e.g., northern midlatitudes and the Arctic). In the latter, large differences between calculated results and ground-based TOST data are found, albeit no satellite data on tropospheric ozone available for a second examination. These overestimates are most likely caused by the enhanced ozone production due to the halogen chemistry in the troposphere, which must be included in future model development if one desires to achieve a precise estimate of ozone loss in this lower altitude region. We should also note that further improvements may be achieved if time-series variations of lower stratospheric and tropospheric temperatures are included in the calculations of ozone trends. Also, the spatial resolution in theoretical calculations of global ozone depletion by the derived equation can be improved to individual latitudes instead of the broad latitude bands if the cosmic ray flux as a function of latitude is used. Given the new identified critical role of $N_2O_5$ in controlling ozone loss from this study, there is a compelling need of precise measurements of this reactive nitrogen species in the atmosphere.

  Finally, we would like to note here that although this study was not intended to provide accurate calculations of global ozone depletion due to its inclusion of simplified assumptions, the parameter-free CRE analytical equation has given theoretical results in satisfactory agreements with observations in the global stratosphere, ranging from the Antarctic, the Arctic and the tropics to both northern and southern midlatitudes. This achievement renders us confidence in validating our recent discovery of the tropical ozone hole and in applying the equation to achieve quantitative understanding of global ozone depletion. To our understanding, no previous CCMs have been able to provide a better agreement with observations in the global lower stratosphere including all the polar regions, the tropics and midlatitudes simultaneously and consistently. The results presented in this article have demonstrated that the CRE mechanism can remove the persistent discrepancies between CCMs and observations, particularly in the lower stratosphere. Further research on the CRE mechanism will be of interest.

**SUPPLEMENTARY MATERIAL**

See supplementary material for Fig. S1.

**ACKNOWLEDGEMENTS**

The author is greatly indebted to the Science Teams (WOUDC's TOST, NASA's GOZCARDS and HIRDLS, SPARC Data Initiative, EUMETSAT's ROM SAF, etc.) for making the data used for this study available. This work is supported by the Natural Science and Engineering Research Council of Canada.

**AUTHOR DECLARATIONS**

The author has no conflicts to disclose.

**DATA AVAILABILITY**

The data used for this study were obtained from the following sources: the TOST data (*2*) were obtained from the WMO's World Ozone and Ultraviolet Radiation Data Centre (WOUDC) (https://woudc.org/archive/products/ozone/vertical-ozone-profile/ozonesonde/1.0/tost/); the GOZCARDS data (*6*) were obtained from the NASA EARTHDATA dataset (https://disc.gsfc.nasa.gov/datasets?keywords=GOZCARDS); the zonal mean $N_2O_5$ HIRDLS satellite data (*83*) were obtained the SPARC Data Initiative (https://www.sparc-climate.org/data-centre/data-access/sparc-data-initiative/); the altitude profiles and DET cross sections of ODSs were obtained from refs. (*35, 84*) and refs. (*11, 13-15, 17*), respectively; the RO lower stratospheric temperature satellite datasets were obtained from the ROM SAF (https://www.romsaf.org/product_archive.php); the cosmic ray flux data were obtained from ref. (*65*); tropospheric data of ODSs were obtained from the 2021 IPCC AR6 Report.

**References:**

1. Q.-B. Lu, Observation of Large and All-Season Ozone Losses over the Tropics. *AIP Advances* **12**, 075006 (2022).
2. G. Liu *et al.*, A global tropospheric ozone climatology from trajectory-mapped ozone soundings. *Atmos. Chem. Phys.* **13**, 10659-10675 (2013).
3. J. Liu *et al.*, A global ozone climatology from ozone soundings via trajectory mapping: a stratospheric perspective. *Atmos. Chem. Phys.* **13**, 11441-11464 (2013).
4. Q.-B. Lu, Response to "Comment on 'Observation of large and all-season ozone losses over the tropics'" [AIP Adv. 12, 075006 (2022)]. *AIP Advances* **12**, 129101 (2022).
5. W. J. Randel *et al.*, Trends in the Vertical Distribution of Ozone. *Science* **285**, 1689-1692 (1999).

6. L. Froidevaux *et al.*, Global OZone Chemistry And Related trace gas Data records for the Stratosphere (GOZCARDS): methodology and sample results with a focus on HCl, H$_2$O, and O$_3$. *Atmospheric Chemistry and Physics* **15**, 10471-10507 (2015).
7. SPARC/IO3C/GAW, 2019: SPARC/IO3C/GAW Report on Long-term Ozone Trends and Uncertainties in the Stratosphere. I. Petropavlovskikh, S. Godin-Beekmann, D. Hubert, R. Damadeo, B. Hassler, V. Sofieva (Eds.), SPARC Report No. 9, GAW Report No. 241, WCRP-17/2018, doi: 10.17874/f899e57a20b, available at https://www.sparc-climate.org/publications/sparc-reports/sparc-report-no-9/ .
8. U. Langematz, Stratospheric ozone: down and up through the anthropocene. *ChemTexts* **5**, 8 (2019).
9. J. Bloch, J. G. Mihaychuk, H. M. van Driel, Electron Photoinjection from Silicon to Ultrathin SiO$_2$ Films via Ambient Oxygen. *Physical Review Letters* **77**, 920-923 (1996).
10. N. Shamir, J. G. Mihaychuk, H. M. van Driel, H. J. Kreuzer, Universal Mechanism for Gas Adsorption and Electron Trapping on Oxidized Silicon. *Physical Review Letters* **82**, 359-361 (1999).
11. Q.-B. Lu, T. E. Madey, Giant enhancement of electron-induced dissociation of chlorofluorocarbons coadsorbed with water or ammonia ices: Implications for atmospheric ozone depletion. *Journal of Chemical Physics* **111**, 2861-2864 (1999).
12. Q.-B. Lu, T. E. Madey, Negative-ion enhancements in electron-stimulated desorption of CF2Cl2 coadsorbed with nonpolar and polar gases on Ru(0001). *Physical Review Letters* **82**, 4122-4125 (1999).
13. Q.-B. Lu, L. Sanche, Enhanced dissociative electron attachment to CF2Cl2 by transfer of electrons in precursors to the solvated state in water and ammonia ice. *Physical Review B* **63**, 153403 (2001).
14. Q.-B. Lu, L. Sanche, Effects of cosmic rays on atmospheric chlorofluorocarbon dissociation and ozone depletion. *Physical Review Letters* **87**, 078501 (2001).
15. Q.-B. Lu, L. Sanche, Large enhancement in dissociative electron attachment to HCl adsorbed on H2O ice via transfer of presolvated electrons. *Journal of Chemical Physics* **115**, 5711-5713 (2001).
16. Q.-B. Lu, Cosmic-ray-driven electron-induced reactions of halogenated molecules adsorbed on ice surfaces: Implications for atmospheric ozone depletion and global climate change. *Physics Reports-Review Section of Physics Letters* **487**, 141-167 (2010).
17. Q.-B. Lu, *New Theories and Predictions on the Ozone Hole and Climate Change*. (World Scientific, New Jersey, 2015), pp. 1-285 pages.
18. M. Bertin *et al.*, Reactivity of water-electron complexes on crystalline ice surfaces. *Faraday Discussions* **141**, 293-307 (2009).
19. J. Staehler, C. Gahl, M. Wolf, Dynamics and Reactivity of Trapped Electrons on Supported Ice Crystallites. *Accounts of Chemical Research* **45**, 131-138 (2012).
20. I. I. Fabrikant, Dissociative electron attachment to halogen molecules: Angular distributions and nonlocal effects. *Physical Review A* **94**, 052707 (2016).
21. I. I. Fabrikant, Electron attachment to molecules in a cluster environment: suppression and enhancement effects. *The European Physical Journal D* **72**, 96 (2018).
22. M. J. Molina, F. S. Rowland, Stratospheric sink for chlorofluoromethanes - chlorine atomic-catalysed destruction of ozone. *Nature* **249**, 810-812 (1974).

23. R. Stolarski, R. Cicerone, Stratospheric chlorine - possible sink for ozone. *Canadian Journal of Chemistry* **52**, 1610-1615 (1974).
24. P. J. Crutzen, The influence of nitrogen oxides on the atmospheric ozone content. *Quarterly Journal of the Royal Meteorological Society* **96**, 320-325 (1970).
25. F. S. Rowland, Stratospheric Ozone Depletion by Chlorofluorocarbons (Nobel Lecture). *Angewandte Chemie International Edition in English* **35**, 1786-1798 (1996).
26. J. C. Farman, B. G. Gardiner, J. D. Shanklin, Large losses of total ozone in Antarctica reveal seasonal ClOx/NOx interaction. *Nature* **315**, 207-210 (1985).
27. S. Solomon, R. R. Garcia, F. S. Rowland, D. J. Wuebbles, On the depletion of Antarctic ozone. *Nature* **321**, 755-758 (1986).
28. M. A. Tolbert, M. J. Rossi, R. Malhotra, D. M. Golden, Reaction of Chlorine Nitrate with Hydrogen Chloride and Water at Antarctic Stratospheric Temperatures. *Science* **238**, 1258-1260 (1987).
29. M. A. Tolbert, M. J. Rossi, D. M. Golden, Antarctic Ozone Depletion Chemistry: Reactions of $N_2O_5$ with $H_2O$ and HCl on Ice Surfaces. *Science* **240**, 1018-1021 (1988).
30. P. J. Crutzen, F. Arnold, Nitric acid cloud formation in the cold Antarctic stratosphere: a major cause for the springtime 'ozone hole'. *Nature* **324**, 651-655 (1986).
31. M. B. McElroy, R. J. Salawitch, S. C. Wofsy, Antarctic $O_3$: Chemical mechanisms for the spring decrease. *Geophysical Research Letters* **13**, 1296-1299 (1986).
32. L. T. Molina, M. J. Molina, Production of chlorine oxide ($Cl_2O_2$) from the self-reaction of the chlorine oxide (ClO) radical. *The Journal of Physical Chemistry* **91**, 433-436 (1987).
33. Y. L. Yung, M. Allen, D. Crisp, R. W. Zurek, S. P. Sander, Spatial Variation of Ozone Depletion Rates in the Springtime Antarctic Polar Vortex. *Science* **248**, 721-724 (1990).
34. J. G. Anderson, D. W. Toohey, W. H. Brune, Free-radicals within the antarctic vortex - the role of cfcs in antarctic ozone loss. *Science* **251**, 39-46 (1991).
35. G. Brasseur, J. J. Orlando, G. S. Tyndall, National Center for Atmospheric Research (U.S.), *Atmospheric Chemistry and Global Change*. Topics in environmental chemistry (Oxford University Press, New York, 1999), pp. xviii, 654 p.
36. B. T. Jobson, Niki, H., Yokouchi, Y., Bottenheim, J., Hopper, F., and Leaitch, R., Measurements of C2-C6 hydrocarbons during the Polar Sunrise1992 Experiment: Evidence for Cl atom and Br atom chemistry. *Journal of Geophysical Research: Atmospheres* **99**, 25355-25368 (1994).
37. B. J. Finlayson-Pitts, The Tropospheric Chemistry of Sea Salt: A Molecular-Level View of the Chemistry of NaCl and NaBr. *Chemical Reviews* **103**, 4801-4822 (2003).
38. B. J. Finlayson-Pitts, Halogens in the troposphere. *Anal Chem* **82**, 770-776 (2010).
39. W. R. Simpson, S. S. Brown, A. Saiz-Lopez, J. A. Thornton, R. von Glasow, Tropospheric Halogen Chemistry: Sources, Cycling, and Impacts. *Chemical Reviews* **115**, 4035-4062 (2015).
40. O. W. Wingenter *et al.*, Tropospheric hydroxyl and atomic chlorine concentrations, and mixing timescales determined from hydrocarbon and halocarbon measurements made over the Southern Ocean. *Journal of Geophysical Research: Atmospheres* **104**, 21819-21828 (1999).
41. W. Hickam, D. Berg, Negative ion formation and electric breakdown in some halogenated gases. *Journal of Chemical Physics* **29**, 517-523 (1958).

42. L. Christophorou, Electron-attachment to molecules in dense gases (quasi-liquids). *Chemical Reviews* **76**, 409-423 (1976).
43. E. Illenberger, H. Scheunemann, H. Baumgartel, Negative-ion formation in $CF_2Cl_2$, $CF_3Cl$ and $CFCl_3$ following low-energy (0-10ev) impact with near monoenergetic electrons. *Chemical Physics* **37**, 21-31 (1979).
44. S. Peyerimhoff, R. Buenker, Potential curves for dissociative electron-attachment of $CFCl_3$. *Chemical Physics Letters* **65**, 434-439 (1979).
45. M. Lewerenz, B. Nestmann, P. Bruna, S. Peyerimhoff, The electronic-spectrum, photodecomposition and dissociative electron-attachment of $CF_2Cl_2$ - an abinitio configuration-interaction study. *Journal of Molecular Structure-Theochem* **24**, 329-342 (1985).
46. Q.-B. Lu, A. Bass, L. Sanche, Superinelastic electron transfer: Electron trapping in $H_2O$ ice via the $N_2^-$ resonance. *Physical Review Letters* **88**, (2002).
47. C. Wang, T. Luo, Q.-B. Lu, On the lifetimes and physical nature of incompletely relaxed electrons in liquid water. *Physical Chemistry Chemical Physics* **10**, 4463-4470 (2008).
48. S. Ryu, J. Chang, H. Kwon, S. Kim, Dynamics of solvated electron transfer in thin ice film leading to a large enhancement in photodissociation of $CFCl_3$. *Journal of the American Chemical Society* **128**, 3500-3501 (2006).
49. H. Hotop, M.-W. Ruf, J. Kopyra, T. M. Miller, I. I. Fabrikant, On the relation between the activation energy for electron attachment reactions and the size of their thermal rate coefficients. *The Journal of Chemical Physics* **134**, 064303 (2011).
50. Q.-B. Lu, T. E. Madey, L. Parenteau, F. Weik, L. Sanche, Structural and temperature effects on $Cl^-$ yields in electron-induced dissociation of $CF_2Cl_2$ adsorbed on water ice. *Chemical Physics Letters* **342**, 1-6 (2001).
51. F. Fehsenfeld *et al.*, Ion chemistry of chlorine compounds in troposphere and stratosphere. *Journal of Geophysical Research-Oceans and Atmospheres* **81**, 4454-4460 (1976).
52. D. Smith and N. G. Adams, Elementary plasma reactions of environmental interest. Top. Curr. Chem. 89, 1-43 (1980).
53. D. G. Torr, The Photochemistry of the Upper Atmosphere in J. S. Levine, Ed. The Photochemistry of atmospheres: Earth, the other planets, and comets (Academic Press, 1985) Chapt. 5, pp. 165-278.
54. K. Oum, M. Lakin, D. DeHaan, T. Brauers, B. Finlayson-Pitts, Formation of molecular chlorine from the photolysis of ozone and aqueous sea-salt particles. *Science* **279**, 74-77 (1998).
55. K. Oum, M. Lakin, B. Finlayson-Pitts, Bromine activation in the troposphere by the dark reaction of O-3 with seawater ice. *Geophysical Research Letters* **25**, 3923-3926 (1998).
56. E. M. Knipping *et al.*, Experiments and Simulations of Ion-Enhanced Interfacial Chemistry on Aqueous NaCl Aerosols. *Science* **288**, 301-306 (2000).
57. H. D. Osthoff *et al.*, High levels of nitryl chloride in the polluted subtropical marine boundary layer. *Nature Geoscience* **1**, 324-328 (2008).
58. J. M. Roberts, H. D. Osthoff, S. S. Brown, A. R. Ravishankara, $N_2O_5$ Oxidizes Chloride to $Cl_2$ in Acidic Atmospheric Aerosol. *Science* **321**, 1059-1059 (2008).

59. J. D. Raff *et al.*, Chlorine activation indoors and outdoors via surface-mediated reactions of nitrogen oxides with hydrogen chloride. *Proceedings of the National Academy of Sciences* **106**, 13647-13654 (2009).
60. A. D. Hammerich, B. J. Finlayson-Pitts, R. B. Gerber, Mechanism for formation of atmospheric Cl atom precursors in the reaction of dinitrogen oxides with HCl/Cl⁻ on aqueous films. *Physical Chemistry Chemical Physics* **17**, 19360-19370 (2015).
61. E. E. McDuffie *et al.*, ClNO$_2$ Yields From Aircraft Measurements During the 2015 WINTER Campaign and Critical Evaluation of the Current Parameterization. *Journal of Geophysical Research: Atmospheres* **123**, 12,994-913,015 (2018).
62. Q.-B. Lu, Cosmic-ray-driven reaction and greenhouse effect of halogenated molecules: culprits for atmospheric ozone depletion and global climate change. *International Journal of Modern Physics B* **27**, 1350073 (2013).
63. Q.-B. Lu, Fingerprints of the cosmic ray driven mechanism of the ozone hole. *AIP Advances* **11**, 115307 (2021).
64. WMO, *Scientific assessment of ozone depletion: 2006*, Global Ozone Research and Monitoring Project - Report No. 50, (Geneva, Switzerland, 2007).
65. G. A. Bazilevskaya, M. B. Krainev, V. S. Makhmutov, Effects of cosmic rays on the Earth's environment. *Journal of Atmospheric and Solar-Terrestrial Physics* **62**, 1577-1586 (2000).
66. J. Kiefer, *Biological radiation effects*. (Springer-Verlag, Berlin ; New York, 1990), pp. 97.
67. Q.-B. Lu, Effects and applications of ultrashort-lived prehydrated electrons in radiation biology and radiotherapy of cancer. *Mutation Research-Reviews in Mutation Research* **704**, 190-199 (2010).
68. I. G. Usoskin, O. G. Gladysheva, G. A. Kovaltsov, Cosmic ray-induced ionization in the atmosphere: spatial and temporal changes. *Journal of Atmospheric and Solar-Terrestrial Physics* **66**, 1791-1796 (2004).
69. A. Adriani, T. Deshler, G. D. Donfrancesco, G. P. Gobbi, Polar stratospheric clouds and volcanic aerosol during spring 1992 over McMurdo Station, Antarctica: Lidar and particle counter comparisons. *Journal of Geophysical Research: Atmospheres* **100**, 25877-25897 (1995).
70. G. P. Gobbi, Lidar estimation of stratospheric aerosol properties: Surface, volume, and extinction to backscatter ratio. *Journal of Geophysical Research: Atmospheres* **100**, 11219-11235 (1995).
71. M. C. Pitts, L. R. Poole, R. Gonzalez, Polar stratospheric cloud climatology based on CALIPSO spaceborne lidar measurements from 2006 to 2017. *Atmos. Chem. Phys.* **18**, 10881-10913 (2018).
72. Q.-B. Lu, Major Contribution of Halogenated Greenhouse Gases to Global Surface Temperature Change. *Atmosphere* **13**, 1419 (2022).
73. K. R. Wilson, A. M. Prophet, M. D. Willis, A Kinetic Model for Predicting Trace Gas Uptake and Reaction. *The Journal of Physical Chemistry A* **126**, 7291-7308 (2022).
74. Q.-B. Lu, L. Sanche, Enhancements in dissociative electron attachment to CF4, chlorofluorocarbons and hydrochlorofluorocarbons adsorbed on H2O ice. *Journal of Chemical Physics* **120**, 2434-2438 (2004).

75. F. Baletto, C. Cavazzoni, S. Scandolo, Surface Trapped Excess Electrons on Ice. *Physical Review Letters* **95**, 176801 (2005).
76. U. Bovensiepen *et al.*, A Dynamic Landscape from Femtoseconds to Minutes for Excess Electrons at Ice−Metal Interfaces. *The Journal of Physical Chemistry C* **113**, 979-988 (2009).
77. I. A. Mironova *et al.*, Energetic Particle Influence on the Earth's Atmosphere. *Space Science Reviews* **194**, 1-96 (2015).
78. D. R. Bates, Recombination of small ions in the troposphere and lower stratosphere. *Planetary and Space Science* **30**, 1275-1282 (1982).
79. D. Smith and N. G. Adams, Ionic recombination in the stratosphere. *Geophysical Research Letters* **9**, 1085-1087 (1982).
80. A. A. Viggiano and F. Arnold, Ion Chemistry and Composition of the Atmosphere. In H. Volland Ed., *Handbook of atmospheric electrodynamics* (CRC Press, Boca Raton, 1995) Chapt 1.
81. J. A. Davidson *et al.*, Rate constants for the reactions of $O_2^+$, $NO_2^+$, $NO^+$, $H_3O^+$, $CO_3^-$, $NO_2^-$, and halide ions with $N_2O_5$ at 300 K. *The Journal of Chemical Physics* **68**, 2085-2087 (1978).
82. Gille, John and Gray, Lesley J. (2013), HIRDLS/Aura Level 3 Dinitrogen Pentoxide ($N_2O_5$) 1deg Lat Zonal Fourier Coefficients V007, Greenbelt, MD, USA, Goddard Earth Sciences Data and Information Services Center (GES DISC), Accessed: [Data Access Date], 10.5067/Aura/HIRDLS/DATA310.
83. M. I. Hegglin *et al.*, Overview and update of the SPARC Data Initiative: comparison of stratospheric composition measurements from satellite limb sounders. *Earth Syst. Sci. Data* **13**, 1855-1903 (2021).
84. R. Zander *et al.*, The 1985 chlorine and fluorine inventories in the stratosphere based on ATMOS observations at 30° north latitude. *Journal of Atmospheric Chemistry* **15**, 171-186 (1992).
85. S. M. Frith *et al.*, Recent changes in total column ozone based on the SBUV Version 8.6 Merged Ozone Data Set. *Journal of Geophysical Research: Atmospheres* **119**, 9735-9751 (2014).
86. L. Froidevaux *et al.*, Validation of Aura Microwave Limb Sounder HCl measurements. *Journal of Geophysical Research: Atmospheres* **113**, (2008).
87. J. F. Noxon, Stratospheric $NO_2$: 2. Global behavior. *Journal of Geophysical Research: Oceans* **84**, 5067-5076 (1979).
88. S. Manabe, R. T. Wetherald, Thermal Equilibrium of the Atmosphere with a Given Distribution of Relative Humidity. *Journal of Atmospheric Sciences* **24**, 241-259 (1967).
89. K. P. Shine *et al.*, A comparison of model-simulated trends in stratospheric temperatures. *Quarterly Journal of the Royal Meteorological Society* **129**, 1565-1588 (2003).
90. SPARC, Report No. 5: CCMVal Report on the Evaluation of Chemistry-Climate Models. V. Eyring, T. Shepherd and D. Waugh (Eds.), (2010), available at www.sparc-climate.org/publications/sparc-reports/. Chapts 4 and 9.

#